\documentclass[preprint,
nobibnotes,
eqsecnum,
]{revtex4}

\usepackage[dvipsnames]{xcolor}
\usepackage{epsfig}
\usepackage{amsfonts}
\usepackage{amsmath}
\usepackage{wasysym}
\usepackage{amssymb}
\usepackage{bm}
\usepackage{float}
\usepackage{enumitem}
\usepackage{rotating}

\usepackage{pdflscape}
\usepackage{graphicx}
\usepackage{ulem}

\usepackage[utf8]{inputenc}
\usepackage{bbm} 				
\usepackage{epstopdf}				
\usepackage{verbatim}    			
\usepackage{sidecap}                
\usepackage{indentfirst}

\newcommand{\be}{\begin{equation}}
\newcommand{\ee}{\end{equation}}
\newcommand{\bea}{\begin{eqnarray}}
\newcommand{\eea}{\end{eqnarray}}

\definecolor{armygreen}{rgb}{0.0, 0.5, 0.0}

\begin{document}

\title{Circulation Statistics in Rayleigh-Bénard Convection}

\author{
Giovanni Saisse$^{1}$,
 Roshan J. Samuel$^{2}$,
Luca Moriconi$^{1}$,
Jörg Schumacher$^{2}$,
and
Katepalli R. Sreenivasan$^{3,4}${\footnote{Corresponding author: katepalli.sreenivasan@nyu.edu}}}
\affiliation{\\}
\affiliation{$^{1}$Instituto de F\'{i}sica, Universidade Federal do Rio de Janeiro, 21941-909, Rio de Janeiro, RJ, Brazil}
\affiliation{$^{2}$Institute of Thermodynamics and Fluid Mechanics, Technische Universit\"at Ilmenau, P.O.Box 100565,
D-98684 Ilmenau, Germany}
\affiliation{$^{3}$Department of Mechanical and Aerospace Engineering, New York University, New York 11201 USA}
\affiliation{$^{4}$Department of Physics and Courant Institute of Mathematical Sciences,
New York University, New York 10012, USA}


\begin{abstract}
Important statistical properties of velocity circulation in homogeneous isotropic turbulence (HIT) have been unveiled in recent years, raising the question of whether they persist, or are modified, in other classes of turbulent flows. Motivated by the dominant role of small-scale structures in the circulation fluctuations of HIT, we investigate their relevance in direct numerical simulations of Rayleigh–Bénard convection at a Rayleigh number of $10^9$ and Prandtl number of $O(1)$. Within the thermal boundary layer (TBL), the distribution of elementary vortices is found to be strongly correlated with the temperature field, while the statistics of their core aspect ratios is significantly altered.~Additionally, the probability distribution functions of circulation, computed for planar contours which are parallel to the walls, display salient features closely akin to those observed in HIT, with the {\it Area Rule} (a connection between circulation statistics and minimal surfaces) remaining particularly well satisfied---except for a possible transitional region at a distance of a few TBL thicknesses. Away from the TBL, as expected, the overall statistical behavior of these structures likewise resembles that of HIT, where the intermittent spatial distribution of small vortex tubes is instead determined by the energy dissipation field.

\end{abstract}


\maketitle

\section{Introduction}


Turbulent flows are characterized by complex spatiotemporal patterns of kinetic energy dissipation and vorticity across a broad range of scales \cite{Tennekes-Lumley, frisch, sreeniRMP}. As experiments and simulations in homogeneous and isotropic turbulence (HIT) have clearly indicated, vorticity is typically organized in the form of coherent structures, such as viscosity-regularized vortex layers and vortex tubes \cite{orszag_etal, farge_etal, kaneda_etal,
kaneda_etal2, sreeniRMP, ishihara_etal, jorg_etal, afonso_etal, afonso2_etal}. 
In fact, they provide a kinematic context that is expected to arise from solutions of the 
fluid dynamical equations.

A convenient and unifying way to probe statistical properties of turbulent flows, with emphasis on the existence of vortex structures, is provided by the velocity circulation
\begin{equation}
\Gamma [C] = \oint_C dx_i \, v_i(x) \ , \  \label{circ}
\end{equation}
evaluated over arbitrary oriented closed contours $C$. 
Due to the Galilean invariance of $\Gamma[C]$, circulation fluctuations tend to suppress the {\it sweep effect} of large eddies on smaller ones. Similar to the role played by velocity structure functions in the statistical theory of turbulence \cite{frisch}, circulation can be used in investigating the multiscale structure of turbulent flows. The relevance of this methodological remark has long been emphasized in the literature \cite{migdal}, but only with the advent of high-performance computing platforms in the study of HIT \cite{Iyer_etal,Iyer_etal2} has circulation been brought into recent spotlight.

The discovery of bifractality of circulation \cite{Iyer_etal} and the connection between circulation statistics and geometrical features of contour loops \cite{Iyer_etal2} prompted a major research effort toward their theoretical understanding. A result of this effort is that these and other related circulation phenomena can now be explained quantitatively, and also predicted by modeling HIT as a system of small-sized elementary vortices with suitably defined circulation and spatial statistical distributions \cite{apol_etal, bounded_measures, mori_etal,mori_pereira,mori_etal2,iyer_mori,moriconi_pereira_PTRS,optimal_surfaces,lima_etal}. This kinetic theory of circulation fluctuations is commonly referred to as the vortex gas model (VGM) of circulation statistics.

On another research front, numerical and experimental studies of the statistical behavior of circulation have been carried out in alternative phenomenological settings, such as two-dimensional turbulence \cite{muller_krs,xie_etal}, wall-bounded flows \cite{mug-thor, duan_etal,Song_Xu_2026}, and quantum turbulence \cite{muller_etal, polanco_etal, muller_etal2, muller-krst2}. Remarkably, these systems share a number of statistical features of circulation that are similar to those observed in HIT.

Motivated by these developments, we address in this work the statistics of circulation in turbulent Rayleigh--Bénard convection at moderately high Rayleigh numbers and Prandtl number of $O(1)$, a regime of central importance in atmospheric and environmental convection as well as in laboratory experiments \cite{ChillaSchumacher2012}. This flow is particularly interesting because thermal plumes organize and drive the coherent vortical motions, whose statistical organization remains largely unexplored. These vortical motions are likely to be strongly coupled to the temperature field within the thermal boundary layer, which is known to have noteworthy self-similar properties \cite{shevkar_etal}.

Our main objective is therefore to characterize circulation fluctuations and identify their underlying elementary vortices, assessing the extent to which the statistical phenomenology established for HIT extends to buoyancy-driven turbulent flows. In particular, we investigate whether the circulation statistics and the associated flow structures display signatures that are consistent with 
the VGM description of HIT.

This paper is organized as follows. In Sec. II, we briefly recall the governing equations of the simulated flow, define the observables of interest and outline our methodology. In Sec. III, we discuss the distribution and morphology of the elementary vortices within and outside the thermal boundary layer. Next, in Secs. IV and V, we focus on circulation fluctuations in these same regions, considering both elementary vortices and contours of different sizes and aspect ratios, with particular emphasis on the relevance of VGM ideas. Finally, in Sec. VI, we summarize our main findings and discuss directions for future research.

\section{Model Setup, Observables, and Methodology}

The Rayleigh--Bénard convection (RBC) problem refers to the 
buoyancy-driven flow of an incompressible fluid confined between two parallel horizontal walls maintained at temperatures that differ by $\Delta T$. Throughout this work, we rely on data (velocity and temperature fields) obtained from the direct numerical simulations (DNS) of the dimensionless Boussinesq equations \cite{verma_book}
\bea
&&\partial_t u_i + u_j \partial_j u_i = -\partial_i p + T\, \delta_{i3} + \sqrt{\frac{Pr}{Ra}} \partial^2 u_i \ , \ 
\\
&&\partial_t T + u_i \partial_i T = \frac{1}{\sqrt{Pr Ra}} \partial^2 T \ , \  \\
&& \partial_i u_i = 0 \ . \ 
\eea
The data correspond to a Rayleigh number $Ra = 10^9$, and a Prandtl number $Pr = 0.7$ \cite{samuel_etal, shevkar_etal}.

The simulation domain is a box with no-slip and constant-temperature conditions at top and bottom walls separated by height $H$, and a lateral extent $L/H  = 4$ in the two-periodically bounded horizontal directions. The spectral mesh has $150 \times 150 \times 96$ elements. The flow field within individual elements are computed using Legendre polynomials of order 9, yielding $10^3$ Gauss-Lobatto collocation points per element. In total, the mesh has $1350 \times 1350 \times 864$ collocation points \cite{samuel_etal}. The circulation statistics are then computed on 2D horizontal cross-sectional planes of $2001 \times 2001$ uniformly spaced points, onto which the velocity and temperature fields are interpolated with spectral accuracy. The height of the thermal boundary layer non-dimensionalized by the height of the cell, $\delta_T/H$, can be estimated from the mean vertical temperature profiles or the temperature fluctuation profiles. Both measures agree well with the relation $\delta_T/H = 1/(2Nu)$, where $Nu$ is the non-dimensionalized convective heat transfer. For the simulation at $Ra = 10^9$ and $Pr = 0.7$ in a Cartesian domain of $L/H =4$, the Nusselt number has been calculated as $Nu = 60.9$, which yields $\delta_T/H = 8.21 \times 10^{-3}$. However, the $\delta_T$ values used to compute circulation statistics is based on the temperature fluctuation profiles, $\delta_{T,\mathrm{rms}} = 7.47 \times 10^{-3}$, which will be denoted as $\delta_T$ henceforth \cite{samuel_etal}.

We are interested in analyzing the statistical properties of circulation, Eq.~(\ref{circ}), for rectangular contours of various sizes and aspect ratios, defined on planes parallel to the horizontal walls. We consider the ``slicing" planes that are located at distances $\delta_T/4$, $\delta_T/2$, $\delta_T$, $4\delta_T$, and $H/2$ from the fixed-temperature walls, thereby probing the flow both within and outside the thermal boundary layer. 

Keeping in mind the VGM description of circulation statistics in HIT \cite{apol_etal, bounded_measures, mori_etal,mori_pereira,mori_etal2,iyer_mori,moriconi_pereira_PTRS,optimal_surfaces}, a major question we address is the identification and characterization of the elementary vortices of the flow and whether they provide the dominant contributions to circulation fluctuations in the slicing planes. While in HIT the distribution of elementary vortices is determined by the energy dissipation field \cite{mori_pereira,moriconi_pereira_PTRS}, we may expect intermittent thermal plumes to change this picture deep within the thermal boundary layer, owing to buoyancy-induced vortex formation through a mechanism closely related to that responsible for dust devils
\cite{Renno1998}.

The two-dimensional cross-sections of the flow field are 
intercepted by the slicing planes. We identify `vortex spots' in these planes by employing the criterion on the swirling strength \cite{Zhou1999}. Specifically, denoting by $u_1(x_1,x_2)$ and $u_2(x_1,x_2)$ the Cartesian components of the velocity field in the slicing plane, we identify $(x_1,x_2)$ as belonging to a vortex spot whenever the two-dimensional velocity-gradient tensor
\be
\mathbb{A}(x_1,x_2) \equiv 
\begin{bmatrix}
\partial_1 u_1 & \partial_2 u_1 \\
\partial_1 u_2 & \partial_2 u_2
\end{bmatrix}
\ee
has complex eigenvalues $\lambda_R \pm i\, \lambda_I$. As has been known since the first applications of this identification method to turbulent boundary layers \cite{Zhou1999, Herpin2010} that spurious structures may arise for a number of reasons, including vortex packing, background shear, insufficient spatial resolution, and other effects \cite{Elsas_Moriconi}. A practical way to remove these artifacts is to retain only those structures satisfying
$
|\lambda_I| \geq \lambda_{th}
$,
where $\lambda_{th}$ is a prescribed threshold. A common choice is to take $\lambda_{th}$ as proportional to the standard deviation $\sigma_\lambda$ of $\lambda_I$ over the region of the flow under investigation. In the present work, we adopt the relatively small threshold
$
\lambda_{th}= \sigma_\lambda/8
$,
which has been used in previous VGM studies \cite{mori_pereira,moriconi_pereira_PTRS}.










\section{Elementary vortices: Distribution and Shape}

We begin by examining how the distribution and morphology of  elementary vortices correlate with the temperature field and the distance from the heated walls. Snapshots for five slicing planes at different heights (and same instant of time) above the hot surface are shown in Fig.~1. It is clear that the closer the planes are to the wall, the more strongly correlated the vortex spots become with the hotter regions of the flow (where the instantaneous temperature is above the local mean). Due to the up-down parity symmetry of the Boussinesq equations, we expect, and indeed verify, that analogous results hold for planes near the colder wall if the roles of hot and cold fluid parcels are interchanged. 

A simple way to quantify the observed vortex--temperature correlations is through the fraction of identified vortex spots located within the hotter regions of the flow (``hot vortices'') and their relative number densities, normalized by the mean local vortex density. These and other statistical quantities are reported in Table~I (performed for 20 independent flow realizations at each plane height).

\begin{figure*}[t]
    \centering
\includegraphics[scale=0.7]{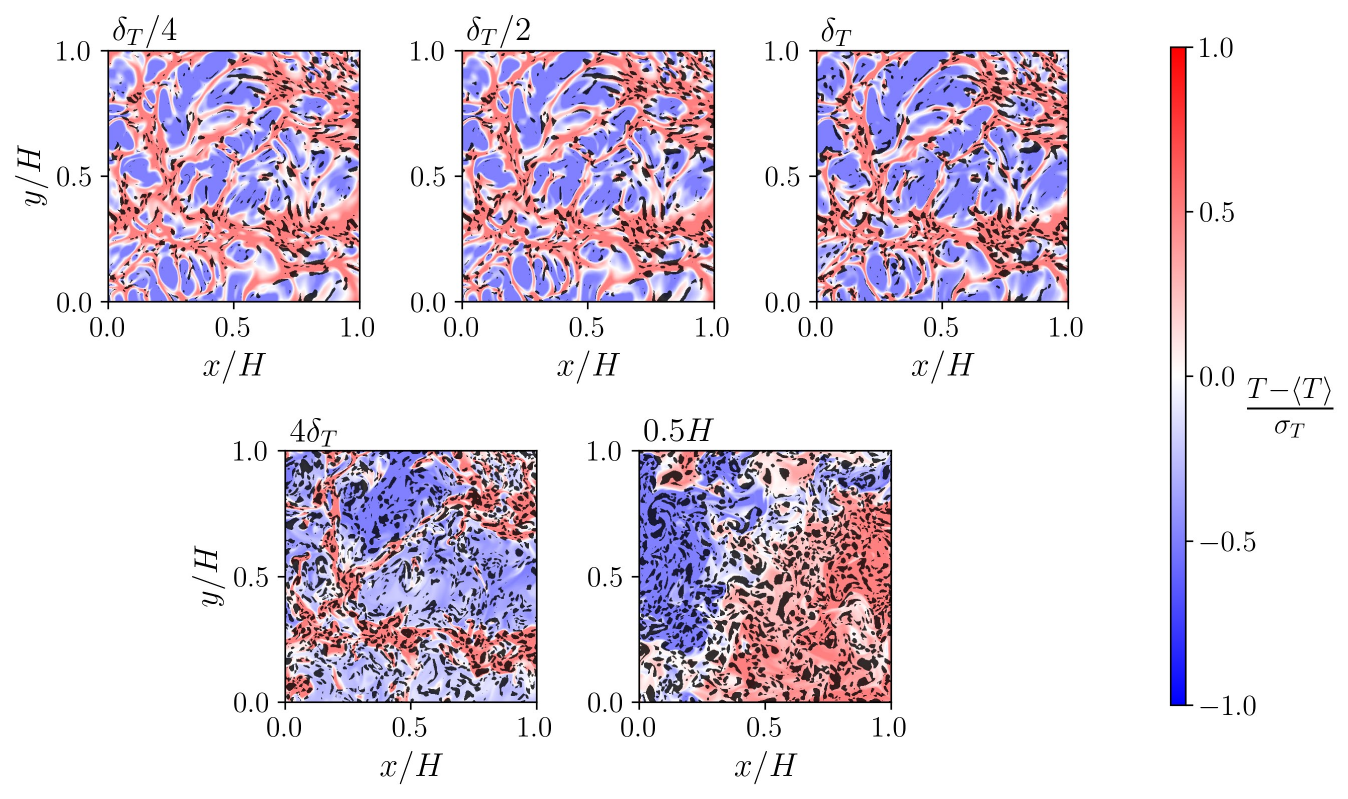}
\caption{Overlay of vortex cross-sections (shown as dark spots) on the temperature field.~The displayed regions represents 1/16 of the total area of each slicing plane.}
        \end{figure*}

\begin{table}[h]
\centering
\begin{tabular}{lcccc}
\hline\hline
Height &-- Hot Vortices (\%) &-- Hot Area (\%) &-- Rel. Density &-- Mean Spot Diam.\\
\hline
$\delta_T/4$ & 67.6 & 55.7 &1.21 & 7.07 \\
$\delta_T/2$ & 65.5 & 53.5 &1.22 & 8.33 \\
$\delta_T$   & 62.1 & 47.1 &1.32 & 8.81 \\
$4\, \delta_T$  & 43.7 & 36.8 &1.19 & 8.60 \\
$H/2$        & 49.1 & 48.1 &1.02 & 10.35 \\
\hline\hline
\end{tabular}
\label{tab:structures}
\caption{Fraction of identified vortex structures lying within the hotter regions of the flow (hot vortices), together with the corresponding area fraction (hot area), the relative number density of hot vortices (defined as the ratio of the second to the third columns), and the mean spot diameters, given in mesh units.}
\end{table}

Two distinct and relevant trends are implied by Table I.~The fraction of hot vortices generally decreases with increasing distance from the hot wall, indicating that the spatial correlation between elementary vortices and hotter fluid regions is strongest deep inside the thermal boundary layer. In contrast, the relative number density of hot vortices reaches its maximum at a height of approximately $\delta_T$, almost up to the top of the thermal boundary layer. This suggests that although a smaller fraction of elementary vortices remains associated with hotter fluid as one moves away from the wall, those that do are most strongly concentrated within the hotter regions near the boundary-layer edge. Beyond this height, the relative number density gradually approaches its bulk value, consistent with a progressive loss of correlation between the temperature and elementary vortices. 

The estimates of the vortex spot diameters listed in Table I also provide important information, as they indicate that the smallest structures are well resolved in the simulation. Their general tendency to increase with height is likely due to the diffusive spreading of the correlated thermal plumes.

    
Another feature that appears to be characteristic of the Rayleigh-Bénard structures is that the vortex spots become increasingly elongated the deeper they lie within the thermal boundary layer (TBL), compared with those found in HIT. To render this observation quantitative, we fit the best ellipse to each identified structure from the second central moments (equivalently, the covariance matrix) of the constituent pixels. Denoting the pixel coordinates by $(x_i,y_i)$, the covariance matrix for a spot with $N$ pixels is
\begin{equation}
\mathbf{C}
=
\frac{1}{N}
\sum_{i=1}^{N}
\begin{bmatrix}
(x_i-\bar{x})^2 &
(x_i-\bar{x})(y_i-\bar{y})
\\
(x_i-\bar{x})(y_i-\bar{y}) &
(y_i-\bar{y})^2
\end{bmatrix},
\label{eq:covariance}
\end{equation}
where
\begin{equation}
\bar{x}=\frac{1}{N}\sum_{i=1}^{N}x_i \ , \
\bar{y}=\frac{1}{N}\sum_{i=1}^{N}y_i \ .\
\end{equation}
The principal axes of the best-fitting ellipse are determined from the eigenvectors of $\mathbf{C}$, while its aspect ratio is given by
\begin{equation}
r=\sqrt{\frac{\lambda_1}{\lambda_2}} \ , \
\end{equation}
where $\lambda_1\ge\lambda_2$ are the eigenvalues of $\mathbf{C}$.
\begin{figure}[h]
\begin{center}
\includegraphics[scale=0.7]{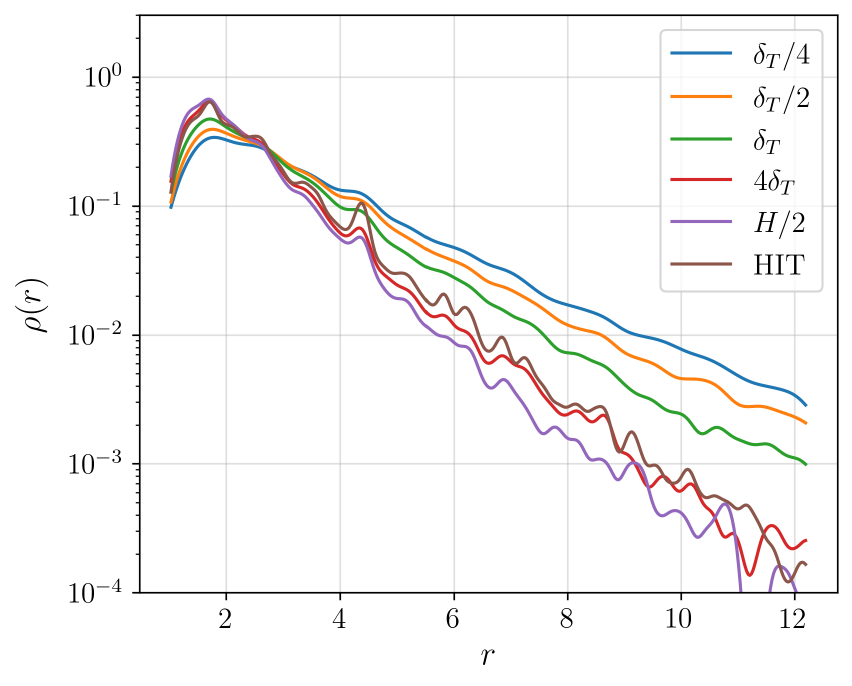}
\caption{{Aspect-ratio PDFs of the identified vortex spots, with the best-fitting ellipses determined from the covariance matrices of the pixelized vortices.}}
\end{center}
\end{figure}

Figure~2 shows, for various heights from the hot wall, the probability density functions (PDFs) of aspect ratios for the identified structures. In all cases, the distributions are characterized by approximately exponential tails.~However, the decay becomes progressively slower as the slicing plane approaches the hotter wall, indicating that the structures indeed become more elongated in the transverse direction within the thermal boundary layer. As the distance from the wall increases beyond $\delta_T$, the PDFs approach the distribution observed in HIT, suggesting a gradual recovery of the homogeneous and isotropic statistics of the elementary vortices in the bulk flow. Our HIT results presented here and in the subsequent comparisons are derived from the Johns Hopkins Turbulence Database (JHTDB) \cite{JHTDB1,JHTDB2}, using the $8192^3$ DNS at $R_\lambda = 1200$.

As an aside, we note that finite-resolution effects introduce spurious correlations in the fine structure of the aspect-ratio PDFs in Fig.~2, for distances greater than about $4\delta_T$. 

Assuming that aspect-ratio fluctuations are mainly driven by the random action of the background strain on polarized vortex spots, we put forward, in Appendix A, a heuristic model based on general ideas from statistical mechanics to account for the exponential tails of the aspect-ratio PDFs.


\section{Elementary vortices and Circulation Statistics}

The snapshots given in Fig.~1 suggest that the vorticity field can be decomposed into a smooth background component and a contribution associated with localized elementary vortices. It is therefore interesting to examine whether the circulation fluctuations around an arbitrary planar contour defined on the sampled planes can be accounted for by the total circulation carried by the individual vortex spots. This hypothesis lies at the heart of the VGM for turbulent circulation statistics in HIT, and here we examine whether it also holds in turbulent RBC.
\begin{figure*}[t]
\centering
\includegraphics[scale=0.7]{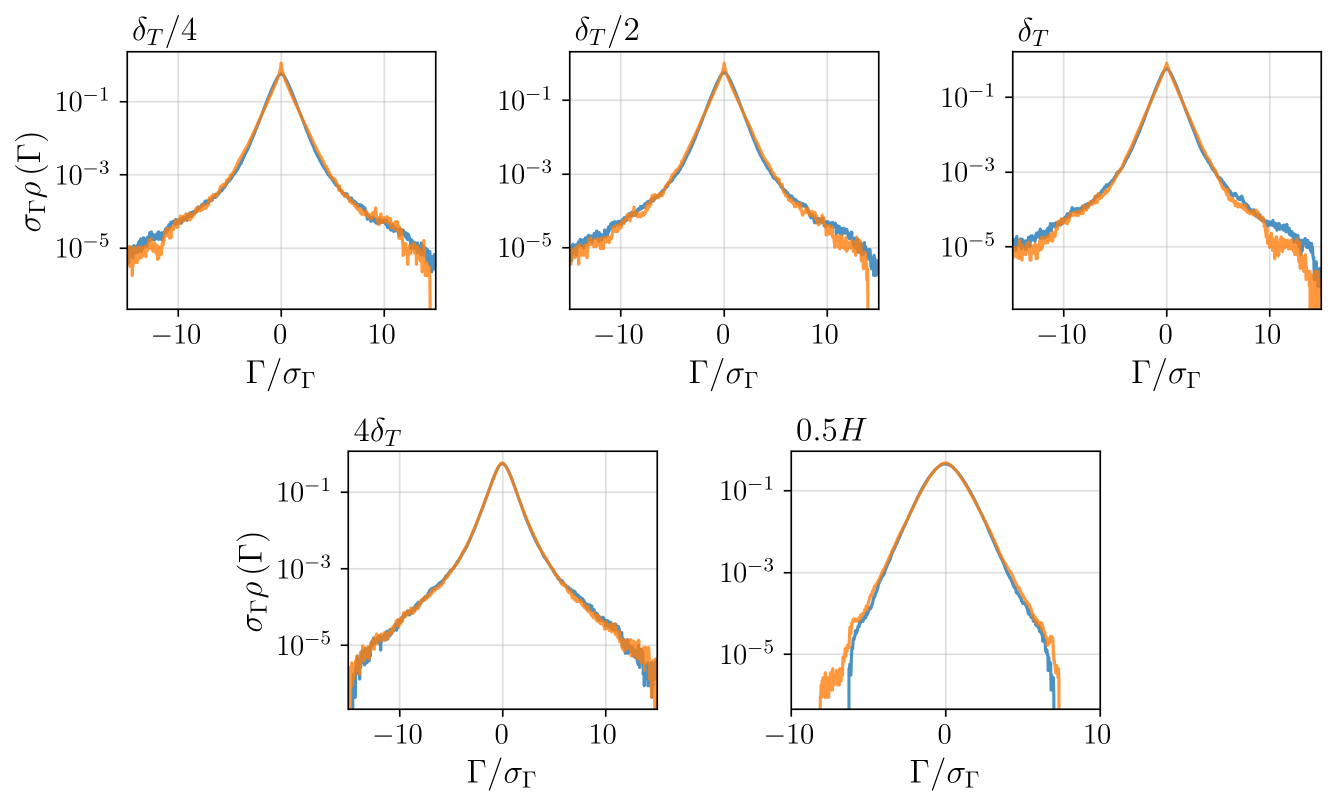}
\caption{Standardized circulation PDFs for square contours of size $50 \times 50$ grid cells, evaluated at different distances from the boundary walls. Blue curves: exact circulation PDFs computed from the vorticity flux; orange curves: circulation PDFs reconstructed from the contributions of vortex spots identified here.}
\end{figure*}

It can be argued that the circulation carried by the identified vortices can be estimated only partially from the flux of the normal vorticity component across the identified vortex spots. The exact circulation would be expected to differ from the circulation measured over the compact vortex 
spots by a multiplicative factor of order unity, determined by the vorticity profile of the elementary vortices \cite{Elsas_Moriconi}. Since this factor is a priori unknown, a pragmatic approach is to compare the standardized PDFs of the exact and the circulation based on vortex spots. Illustrative results are shown in Fig.~3 for circulation PDFs corresponding to square contours of grid size $50 \times 50$, evaluated on each of the five slicing planes. The close agreement between the circulation PDFs thus indicates that the dominant contribution to circulation fluctuations arises from the circulation carried by the vortex spots identified here.

A related question concerns the circulation fluctuations of the elementary vortices themselves. Within the framework of the VGM, the standardized circulation probability density function (PDF) of the elementary vortices in HIT is described by the closed analytical expression \cite{moriconi_pereira_PTRS}
\be
\rho(\Gamma) = \frac{1}{2\pi\sigma}\int_0^{\infty}dY\frac{1}{Y^2}
\exp{\left[-\frac{1}{2\sigma^2}\left(\ln(Y) + \sigma^2\right)^2-\frac{\Gamma^2}{2Y^2}\right]} \ , \ \label{cPDFvs}
\ee
which depends on a single parameter, $\sigma$, obtained from a parabolic fit to the logarithms of the statistical moments, in the form
\begin{equation}
\ln\left(
\frac{\langle|\bar{\Gamma}|^q\rangle}{A_q}
\right) = \frac{1}{2} \sigma^2 \, q^2 + const \times q\ , \
\end{equation}
where $\bar{\Gamma}$ denotes the circulation of the elementary vortices and $A_q$ is the combinatorial factor involving the Euler Gamma function,
\begin{equation}
A_q=
\frac{2^{q/2}}{\sqrt{\pi}}
\Gamma\!\left(\frac{q+1}{2}\right) \ . \
\end{equation}
\begin{figure*}[b]
    \centering
\includegraphics[scale=0.7]{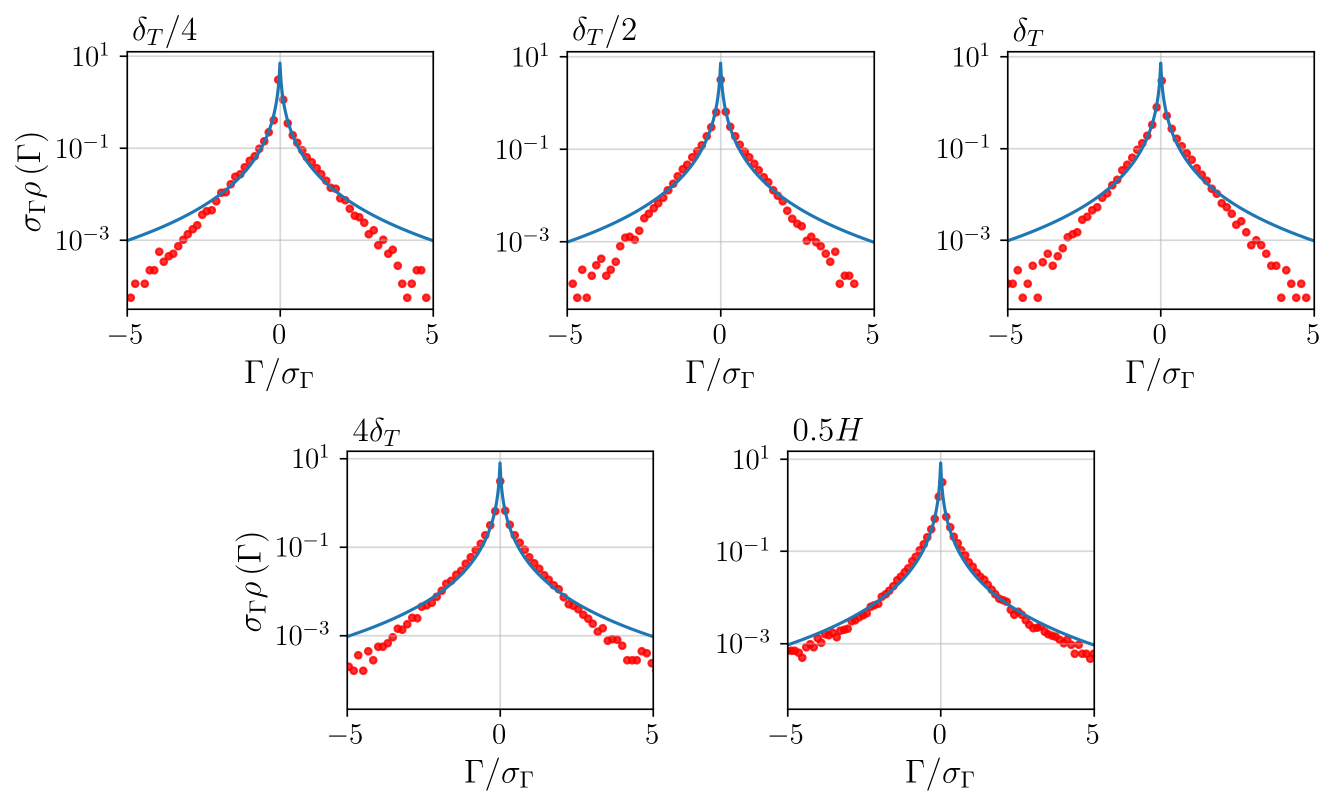}
\caption{Comparison of the analytical circulation PDFs for elementary vortices (HIT), given by Eq.~(\ref{cPDFvs}) (solid blue lines), with the corresponding numerical PDFs in RBC (red symbols).}
\end{figure*} 

\noindent As can be seen from Fig.~4, Eq.~(\ref{cPDFvs}) provides a reasonably good description of the circulation PDFs only sufficiently far from the wall. It is interesting to note that, in HIT, the circulation PDFs of elementary vortices exhibit heavier tails than those observed in the thermal boundary layer of RBC, reflecting the lognormal nature of energy-dissipation fluctuations in HIT \cite{O62,K62}. One might therefore be tempted to conclude that circulation fluctuations around extended domains are less intermittent in the thermal boundary layer. As we show in the next section, however, this is not the case, as deviations from usual Kolmogorov phenomenology are even more pronounced.

We also observe from Figs.~3 and 4 that the standardized circulation PDFs undergo only minor changes within the thermal boundary layer, suggesting an approximately stationary behavior of circulation fluctuations in that region. 



\begin{figure*}[b]
    \centering
\includegraphics[scale=0.7]{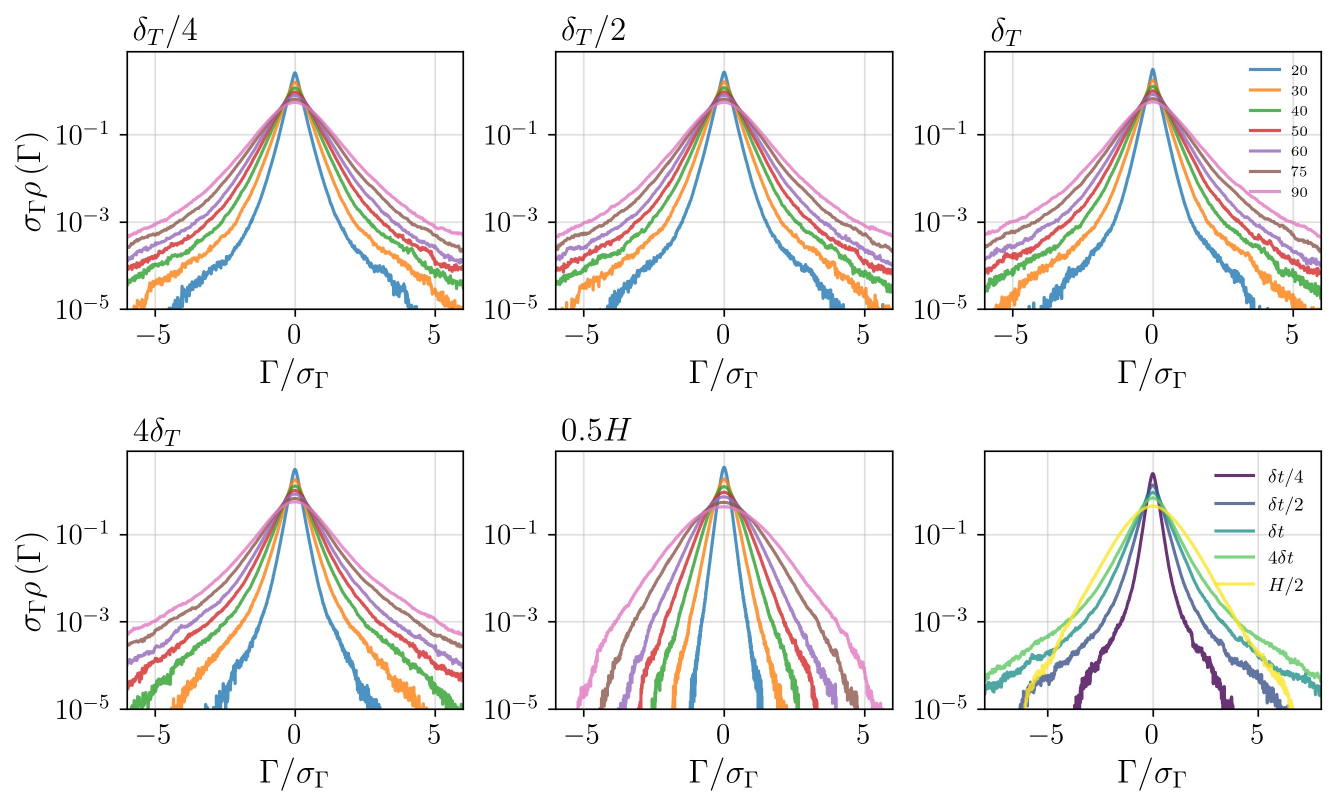}
\caption{Circulation PDFs for square contours of different sizes (side lengths indicated in the upper-right legend) and at different distances from the boundary wall. The PDFs without a height label, shown in the lower-right panel, correspond to square contours of size $50 \times 50$ mesh cells. All PDFs are normalized by the standard deviation of the $H/2$ case, with contour of size $50 \times 50$. }
        \end{figure*}

\begin{figure*}[b]
    \centering
\includegraphics[scale=0.7]{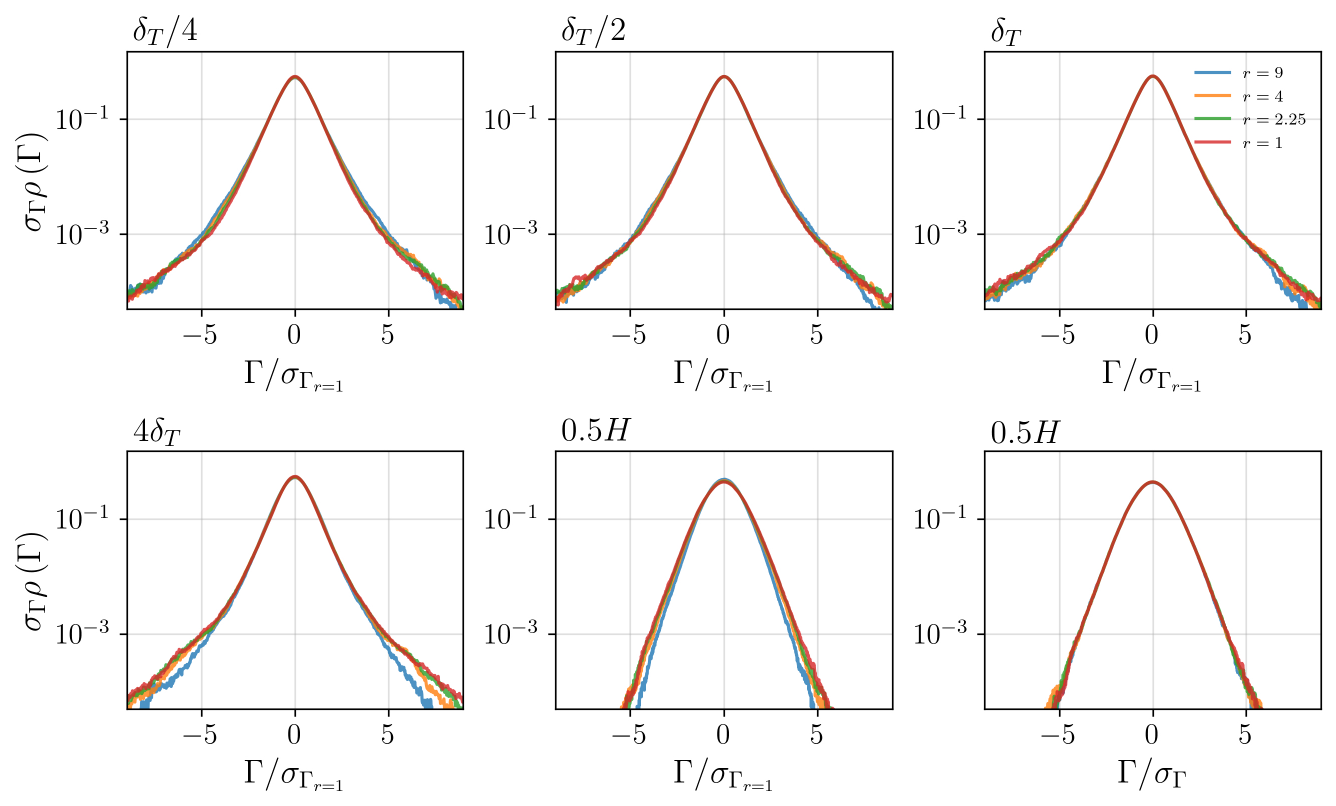}
\caption{Circulation PDFs for all distances from the walls obtained using rectangular loops of fixed area ($60^2$ squared mesh cells) and varying aspect ratio $r$. The distributions are normalized by the standard deviation of the $r=1$ case at each height, except for the standardized PDF shown in the lower right panel.}
        \end{figure*}

\begin{figure}[b]
    \centering
\includegraphics[scale=0.7]{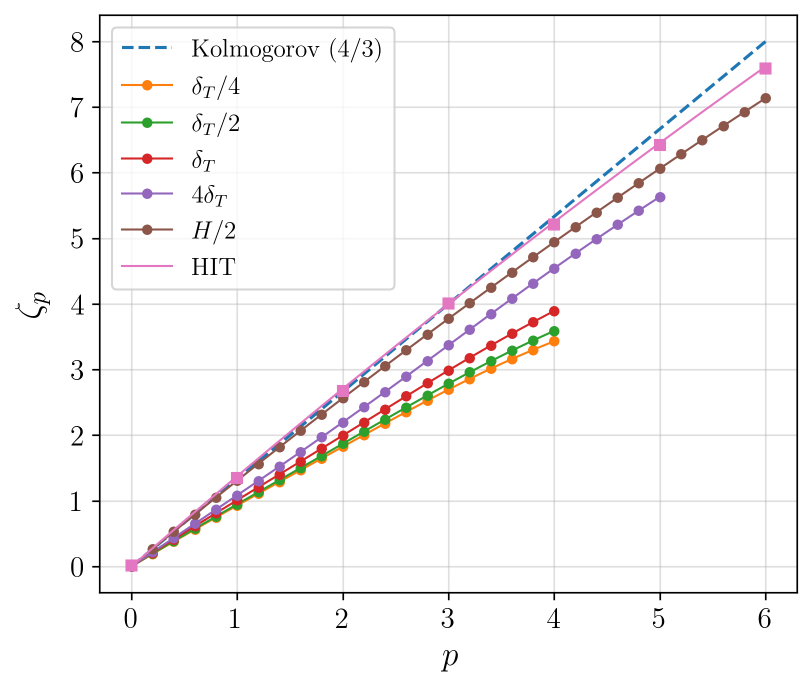}
\caption{In-plane circulation scaling exponents for different distances from the heated walls. The pink solid line and the square symbols correspond to HIT, representing, respectively, the VGM prediction \cite{bounded_measures} and DNS results \cite{Iyer_etal}. The blue dashed line denotes the exact linear scaling predicted by Kolmogorov phenomenology ($\zeta_p = 4p/3$) \cite{Iyer_etal}.}
        \end{figure}

\section{Contour Size Effects and Circulation Scaling}

We now turn our attention to the statistical properties of circulation for contours of different sizes and shapes. Figure~5 shows the circulation PDFs, without normalization by the standard deviations, for square contours ranging in size from $20 \times 20$ to $90 \times 90$ mesh cells. Up to a height of $\delta_T$, the PDFs are remarkably similar. When plotted in standardized form (not shown), they essentially collapse onto a single curve. This provides further support for the approximate self-similar behavior of circulation fluctuations discussed above, which is restricted to the thermal boundary layer. 

The data reveal, in particular, that a contour size of 50 lattice sites already lies near the integral scale of the flow at a distance of $H/2$ from the walls. This is not the case, however, in the thermal boundary layer, where circulation fluctuations are clearly non-Gaussian.

We next examine the influence of contour shapes by considering rectangular loops of fixed area ($50 \times 50$ mesh cells) and a range of aspect ratios, as shown in Fig.~6. Within the thermal boundary layer, the circulation PDFs are found to be nearly independent of the contour aspect ratio. In fact, even without standardization, the distributions collapse to a very good approximation over the entire range of aspect ratios considered. This result indicates that, in the self-similar region close to the walls, circulation statistics are primarily controlled by the enclosed area rather than by the detailed geometry of the integration contour.

Deviations from this behavior become apparent farther away from the walls. At a height of $4\delta_T$, the PDF corresponding to the largest aspect ratio no longer follows the common curve, and standardization only marginally improves the agreement. At $H/2$, the unnormalized PDFs likewise fail to collapse for the most elongated contours. However, once normalized by their respective standard deviations, the distributions recover a reasonably good collapse. This result is consistent with the area rule for circulation statistics, according to which contours enclosing the same area share the same standardized PDF, while their variances retain a residual dependence on the contour geometry outside the inertial range \cite{Iyer_etal2}.

So far, we have collected evidence that the circulation PDFs are approximately self-similar within the thermal boundary layer, in the sense that they differ primarily through their variances when contours of different sizes are considered. A quantitative assessment of this property can be obtained by considering the scaling exponents $\zeta_p$, defined through the circulation moments as
\be
\langle |\Gamma[C]|^p \rangle \sim \ell^{\zeta_p} \ , \
\ee
where $C$ is a closed contour of characteristic linear dimension $\ell$. The circulation scaling exponents were obtained from sets of square contours with side lengths (in mesh units) ranging from $20$ to $90$. They are reported in Fig.~7 for different distances from the wall, up to the highest moment orders that can be estimated with sufficient statistical accuracy.

Figure~7 yields an approximately linear dependence of the scaling exponents on the moment order at low orders, with progressively smaller circulation Hölder exponents,
$
h \equiv d\zeta_p / dp, 
$
as the sampling planes approach the heated wall. The data also suggest weak bifractal scaling at higher moment orders, although larger statistical ensembles are required to establish this conclusively. Significant departures from the HIT exponents are likewise observed at $H/2$, which is similar to the behavior observed in channel flows \cite{duan_etal}. 

\section{Conclusions and Perspectives} 

We have investigated the statistical properties of small-scale vortices and circulation in the thermal boundary layer of turbulent RBC and also in its bulk flow, the latter expected to have features of HIT. A main motivation was to explore whether the key modeling principles underlying the VGM of circulation statistics, originally devised for HIT, could also be realized in RBC.

It turns out that columnar vortices which extend across the entire thermal boundary layer are preferentially embedded within thermal plumes, which transport localized, large-amplitude temperature fluctuations. This constitutes a major distinction from the HIT phenomenology, where the spatial distribution of vortices is instead governed by the dissipation field \cite{mori_pereira,moriconi_pereira_PTRS}. In contrast to HIT, however, where the dissipation field can be reasonably well modeled by the Obukhov--Kolmogorov lognormal model of intermittency \cite{O62,K62} and its various extensions, including that based on the Gaussian multiplicative chaos theory \cite{GMC}, an analogous VGM description of the fluctuating temperature field, and of how it correlates with the vortex distribution in the thermal boundary layer, remains an open and challenging problem.

An interesting point of convergence with HIT comes from the verification that the area law of circulation statistics is equally valid in RBC, for contours lying in planes parallel to the heated walls and within the thermal boundary layer. In this context, the observation that the circulation carried by localized vortices is found to dominate the circulation fluctuations --- a guiding principle of the VGM approach --- appears to be a central phenomenological ingredient in the RBC setting as well.

An intriguing anomalous behavior of circulation fluctuations, at a distance of $4\delta_T$ from the heated walls, deserves further investigation. At this distance from the walls, still well outside the approximately homogeneous and isotropic bulk region, we observe that (i) the fractions of hot vortices and hot area, reported in Table~I, show a non-monotonic behavior and fall below their limiting values 
measured at the height $H/2$, and (ii) the vortex spot diameters show a similar non-monotonic behavior. This non-monotonicity may reflect the existence of a transitional flow regime between the thermal boundary layer and the bulk (although we cannot rule out insufficient statistical convergence). In this transitional region, somewhat surprisingly, the aspect-ratio PDF of vortex spots, is closer to the HIT case, and the area rule is not strictly satisfied: this is in contrast to what one finds separately within the thermal boundary layer and in the bulk (though the normalized PDFs tend to follow the area rule more closely). In general, enhanced ensembles will be necessary to extend the evaluation of the circulation scaling exponents (Fig.~7) to higher-order moments, enabling a more reliable assessment of circulation multifractality.

\acknowledgments

G.S. thanks CAPES for partial support. The work of R.J.S. and J.S. is funded by the European Union (ERC, MesoComp, 101052786). Views and opinions expressed are, however, those of the authors only and do not necessarily reflect those of the European Union or the European Research Council. R.J.S. and J.S. also gratefully acknowledge the Gauss Center for Supercomputing e.V. (https://www.gauss-centre.eu) within the Large Scale Project Nonbou for funding this project by providing computing time on the GCS Supercomputers JUWELS and JUPITER at the J\"ulich Supercomputing Center (https://www.fz-juelich.de/en/ias/jsc). This work was initiated during L.M.'s visit to New York University. Partial support from  CNPq and FAPERJ through Grants $\#$ 311012/2022-1 and $\#$ E-26/200.457/2026 is also acknowledged. K.R.S. thanks the support of New York University for its support of his work.




\appendix

\section{The Distribution of Large Aspect Ratio Fluctuations}

We remark that the exponential tails observed in the aspect-ratio PDFs can be rationalized by simple statistical mechanical arguments.~The circulation, $\Gamma$, and the cross-sectional area, $A$, of a coherent vortex structure are expected to fluctuate only weakly over its lifetime. This behavior is consistent with the approximate conservation of circulation and cross-sectional area, as implied by Kelvin's theorem and incompressibility, respectively.~Under these conditions, the dynamically relevant degree of freedom is the aspect ratio $r$, whose evolution is driven by the fluctuating background strain field. 

Let us suppose, in view of the above considerations, that the circular configuration (\hbox{$r=1$}) of an elliptical isolated vortex minimizes an effective shape function $H(r)$, whereas the surrounding turbulent strain provides a stochastic forcing that continuously deforms the vortex away from this preferred configuration. Since the vortex area is approximately conserved, exchanging the major and minor semi-axes,
\[
r \longleftrightarrow \frac{1}{r} \ , \
\]
merely corresponds to rotating the best-fitting ellipse by $\pi/2$, leaving the physical configuration unchanged. Consequently, the effective shape function must satisfy the symmetry relation
\begin{equation}
H(r)=H(1/r) \ . \
\label{eq:symmetry}
\end{equation}
The lowest-order analytic function satisfying the symmetry relation (\ref{eq:symmetry}) and possessing a minimum at $r=1$ is
\begin{equation}
H(r)
=
\kappa
\left(
r+\frac{1}{r}-2
\right) \ , \
\label{eq:Eeff}
\end{equation}
where $\kappa$ is some positive constant. 
It is clear that for small departures from circularity ($r=1$),
\begin{equation}
H(r)
\simeq
\kappa(r-1)^2 \ , \
\end{equation}
whereas for highly elongated elementary vortices,
\begin{equation}
H(r)
\simeq
\kappa r \ , \
\qquad r\gg1 \ . \ \label{large_r}
\end{equation}
Thus, the shape function is harmonic in the vicinity of the circular configuration while becoming asymptotically linear for large deformations.

As evidenced in HIT, slender elementary vortices tend to cluster into polarized bundles \cite{mori_etal2}, so that their local dynamics is expected to be effectively two-dimensional. Within this picture, which we extend to turbulent RBC \cite{comment2D}, we argue that $H(r)$ can be identified, up to an additive constant, with the kinetic energy of a two-dimensional vortex spot,
\be
E = \frac{1}{2} \int dA \, \psi(x,y)\,\omega(x,y) \ , \  \label{energy1}
\ee
where $\psi(x,y)$ and $\omega(x,y)$ are, respectively, the stream function and vorticity in a plane perpendicular to the clustered vortices. The vorticity field can be decomposed as
\be
\omega(x,y) = \omega^{b}(x,y) + \omega^v(x,y) \ , \ \label{omega}
\ee
where $\omega^b(x,y)$ and $\omega^v(x,y)$ are the vorticity fields associated with the background flow and with the vortex structure, respectively. In an analogous fashion, we write, for the stream function,
\be
\psi(x,y) = \psi^b(x,y) + \psi^v(x,y) \ . \ \label{psi}
\ee
We immediately get, from (\ref{energy1}-{\ref{psi}}), that
\bea
&&\Delta E \equiv  E -  \frac{1}{2} \int dA \, \psi^b(x,y) \omega^b(x,y) = \nonumber \\
&&=\int dA \, \psi^b(x,y) \omega^v(x,y)
+ \frac{1}{2} \int dA \, \psi^v(x,y) \omega^v(x,y) \ , \  \label{delta_energy}
\eea
where we have used $\omega = -\partial^2 \psi$ to derive the cross term.
Expanding $\psi^s(x,y)$, where $s = b$ or $v$, around the vortex center (placed at the origin), we obtain 
\be
\psi^s(x,y)
=
\psi^s_0
+x\,\psi^s_x
+y\,\psi^s_y +\frac12x^2\psi^s_{xx}
+x y\,\psi^s_{xy}
+\frac12y^2\psi^s_{yy}
+\cdots \ . \ \label{expansion}
\ee
To proceed, we consider, as a simplified model, an elliptical vortex spot with semi-major and semi-minor axes of lengths $a$ and $b$ along the $x$ and $y$ directions, respectively, with vorticity
$\omega^v = \Gamma /\pi ab$ inside the elliptical spot and zero outside. Setting $\psi_0^s = 0$ \cite{commentpsi} and invoking reflection symmetry, the linear and bilinear terms in (\ref{expansion}) make no contribution to (\ref{delta_energy}).
The leading shape-dependent correction is therefore
\begin{equation}
\Delta E
=
\frac{\Gamma}{4 \pi ab}
\left[
\left(\psi^b_{xx} + \frac{1}{2} \psi^v_{xx} \right )\int x^2 \, dA
+
\left(\psi^b_{yy} + \frac{1}{2} \psi^v_{yy} \right )\int y^2\,dA
\right] \ . \ 
\end{equation}
Using the second moments of the ellipse,
\begin{equation}
\int x^2\,dA
=
\frac{\pi}{4}a^3b
\ , \
\int y^2\, dA
=
\frac{\pi}{4}ab^3 \ , \
\end{equation}
we obtain
\begin{equation}
\Delta E
=
\frac{\Gamma}{16}
\left[
a^2 \left(\psi^b_{xx} + \frac{1}{2} \psi^v_{xx} \right )
+
b^2 \left(\psi^b_{yy} + \frac{1}{2} \psi^v_{yy} \right )
\right] \ . \ \label{energy}
\end{equation}
Since the surrounding turbulent flow is statistically isotropic, we estimate the second derivatives of the background stream function as
\begin{equation}
\psi^b_{xx}
\sim
\psi^b_{yy}
\sim
\omega_0 \ , \ \label{BG}
\end{equation}
where $\omega_0$ is a typical scale of background vorticity. For the self-induced contribution, on the other hand, we introduce the estimates
\be
\psi^v_{xx} \sim \Gamma /a^2 \ , \ \psi^v_{yy} \sim \Gamma /b^2 \ . \  \label{SI}
\ee
Substituting (\ref{BG}) in (\ref{energy}), the leading shape dependence of the energy then becomes
\begin{equation}
\Delta' E
\propto 
\omega_0 \, \Gamma  (a^2+b^2) \ , \ \label{energy2}
\end{equation}
where 
\be
\Delta' E =  \Delta E -  \frac{\Gamma}{32}
\left(
a^2 \psi^v_{xx} 
+
b^2 \psi^v_{yy} 
\right) \ , \  
\ee
which, by virtue of (\ref{SI}), is indeed a shape independent quantity. It is not difficult to see, furthermore, that 
\begin{equation}
a^2+b^2
\propto
A\left(r+\frac1r\right) \ , \
\end{equation}
where $A \propto ab$ is the  conserved vortex area and $r=a/b$ is the vortex aspect ratio. From (\ref{energy2}), therefore, the shape-dependent energy variations have the same duality-invariant form as (\ref{eq:Eeff}), viz.,
\begin{equation}
\Delta' E =
C
\left(
r+\frac1r
\right),
\end{equation}
where $C\propto \omega_0 \Gamma  A$.

As the fluctuating turbulent strain field plays the role of an effective stochastic forcing acting on the vortex shape, we may assume that the stationary probability distribution of the aspect ratio is given by an effective Boltzmann weight,
\begin{equation}
P(r)\propto
\exp\!\left[-\beta H(r)\right] \ , \ \label{Pr}
\end{equation}
where the ``inverse temperature" $\beta$ parametrizes the variance of aspect ratio fluctuations. In other words, the turbulent background acts as an effective energy reservoir, continuously exchanging energy with individual vortices through the action of the strain field. In this way, we regard the vortex shape as the microstate label of a subsystem weakly coupled to a large bath of turbulent degrees of freedom, providing a physical mechanism for the emergence of an effective Boltzmann distribution for the aspect ratio.

The asymptotically linear behavior (\ref{large_r}) immediately yields
\begin{equation}
P(r)\sim
\exp(-\beta\kappa r) \ , \
\end{equation}
thereby providing a straightforward explanation for the exponential tails observed in the aspect-ratio PDFs.



\bibliography{references_abbrev}

\end{document}